%% file: main.tex
\documentclass[conference]{IEEEtran}
\IEEEoverridecommandlockouts
\usepackage[english]{babel}
\usepackage{amsthm}
\usepackage{titlesec}
\usepackage[ruled,vlined]{algorithm2e}
\usepackage{algcompatible}
\usepackage{graphicx}
\usepackage{float}
\usepackage{caption}
\usepackage{tikz}
\usepackage{changepage}
\usepackage[utf8]{inputenc}
\usepackage{pgfplots} 
\usepackage{pgfgantt}
\usepackage{pdflscape}
\usepackage[export]{adjustbox}
\pgfplotsset{compat=newest} 
\pgfplotsset{plot coordinates/math parser=false}
\pgfplotsset{compat=1.18}
\usepackage[justification=centering]{caption}
\usepackage{pgfplots}
\usetikzlibrary{spy}

\newtheorem{innercustomgeneric}{\customgenericname}
\providecommand{\customgenericname}{}
\newcommand{\newcustomtheorem}[2]{%
  \newenvironment{#1}[1]
  {%
   \renewcommand\customgenericname{#2}%
   \renewcommand\theinnercustomgeneric{##1}%
   \innercustomgeneric
  }
  {\endinnercustomgeneric}
}

\newcustomtheorem{customthm}{Theorem}
\newcustomtheorem{customlemma}{Lemma}

\usepackage{amsmath}
\usepackage{acronym}
\usepackage{amssymb}
\usepackage{mathtools}
\usepackage{url}
\usepackage{graphicx}  
\usepackage{float}  
\input{auxiliary/commands}
\input{auxiliary/acronyms}
\begin{document}

\title{Secure Spatial Signal Design for ISAC in a Cell-Free MIMO Network}

\author{Steven~Rivetti,
Emil~Björnson, 
        Mikael~Skoglund \\
        
       { School of Electrical Engineering and Computer Science (EECS),
        KTH Royal Institute of Technology, Sweden}
        
       \thanks{
This work was supported by the SUCCESS project (FUS21-0026), funded by the Swedish Foundation for Strategic Research.}}

\maketitle

\begin{abstract}
In this paper, we study a cell-free multiple-input multiple-output  network equipped with  integrated sensing and communication (ISAC) access points (APs). The distributed  APs are used to jointly serve the communication needs of user equipments (UEs) while sensing a target, assumed to be an eavesdropper (Eve).
To increase the system's robustness towards said Eve, we develop an ISAC waveform model that includes artificial noise (AN) aimed at degrading the Eve channel quality.
The central processing unit receives the observations from each AP and calculates the optimal precoding and AN covariance matrices by solving a semi-definite relaxation of a  constrained Cramer-Rao bound (CRB) minimization problem. Simulation results highlight an underlying trade-off between sensing and communication performances: in particular, the UEs signal-to-noise and interference ratio and the maximum Eve's signal to noise ratio are directly proportional to the CRB.
Furthermore, the optimal AN covariance matrix is rank-1 and has a peak in the eve's direction, leading to a surprising inverse-proportionality between the UEs-Eve distance and optimal-CRB magnitude.

\end{abstract}


\section{Introduction}

6G networks and other future communication systems are expected to include sensing functionalities as a fundamental component \cite{buzzi2022integrated}. Assuming an efficient sharing of hardware and wireless resources, the communication infrastructure can have low-cost sensing capabilities and the sensing frequency bands can be used for wireless communications. To do that, however, the \ac{ISAC} system's different components, including the transmission waveform, received data post-processing and the \ac{MIMO} beamforming must be carefully co-designed.
The single ISAC \ac{AP} case has been the main focus of earlier studies \cite{mateos2022model}.
%ahmadipour2022information
In reality, however, numerous ISAC \acp{AP} will operate in the same area, frequency range, and period of time, thus creating interference between each other. This encourages these dispersed \acp{AP} to work together to enhance their performance and mitigate the aforementioned interference \cite{behdad2022power}.
Because of the shared usage of the spectrum and the broadcasting nature of wireless transmission, \ac{ISAC} systems' security is facing several challenges.
On the one hand, Rician channels, which are frequently occurring at mmWave frequencies and contain a \ac{LoS} component, are inextricably linked to the sensing channel. This differs from traditional \ac{PLS} research \cite{liu2017enhancing} in
%hamamreh2018classifications 
communication systems with the premise that the legitimate user channels and intercept channels are independently and identically distributed. 
The confidential  information designated to the  \acp{UE} is included in the dual-functional waveform, namely a waveform that jointly serves the communication  \acp{UE} and senses the sensing target \cite{shi2020constrained}, and is therefore vulnerable to being intercepted by the sensing target.
Reasons for implementing such safety measures can stem from various factors. For instance, in scenarios involving vehicle tracking for safety purposes, it is essential to enable vehicle monitoring while preventing unauthorized access to the ongoing communication. Let us consider a communication network deployed in an open-air industrial area, where incoming vehicles deliver goods. The system aims to track the vehicles autonomously, avoiding accidents, and simultaneously safeguarding the confidentiality of the data traffic to prevent industrial espionage.

From the sensing side, a crucial and intriguing trade-off appears: the power is expected to be directed towards the sensing target, however, the useful signal information must be protected from being intercepted by said target, who is identified as a malicious agent.
On the other hand, regardless of the sensing target's actions and intentions, it is important to increase the system's estimation and detection performance. 
In conventional communication systems, security issues are addressed at the higher levels of the protocol stack with cryptographic techniques \cite{melki2019survey}, even though it is worth mentioning that studies on cryptography frequently assume that the physical layer provides a link that is error-free, but in reality, wireless communications are subject to assaults, increasing the risk of information loss.

\subsection{Related work}
In \cite{demirhan2023cell}, the authors consider a cell-free \ac{MIMO} network equipped with \ac{ISAC}  \acp{AP} and devise a transmit waveform co-design strategy aimed at maximizing the \ac{SNR} of the sensing target, here assumed to be a benign agent,  under a minimum \ac{SINR} constraint for the communication users.
The performance of this transmit strategy, in terms of achieved communication \ac{SINR} and sensing \ac{SNR}, are compared against well-established sensing-prioritized and communication-prioritized design strategies, demonstrating the performance gain achieved by the proposed co-design strategy.
On the other hand, in \cite{su2022sensing}, the idea of adding \ac{AN} to the transmitted waveform to improve the  network \ac{PLS} is introduced.
This technique is an alternative to coding-based schemes aimed at achieving the same goal \cite{zhang2010p}.
This work considers a single \ac{AP} with multiple targets, assumed to be eavedroppers (Eves),  and derives the optimal beamforming strategy by solving a weighted optimization problem.
The objective functions bound together by the weighted optimization are the \ac{CRB} onto the targets' angles and the system's secrecy rate, defined as the minimum difference between each user's achievable rate and a sensing rate, defined as $\log_2(1+\ac{SNR})$. 

\subsection{Contributions}
Contrary to the existing literature, this paper investigates how \ac{AN} can be used by distributed \ac{ISAC} \acp{AP}, cooperating to increase the \ac{PLS} of a cell-free \ac{MIMO} network.
Said network is made of $M$ \acp{AP}  that serve multiple communication users and simultaneously sensing a target, assumed to be an Eve, using the same signal. 
In addition to its monostatic observation, every \ac{AP} receives the reflected echos from all the other \acp{AP} and sends them to a \ac{CPU} trough a back-haul link here assumed error-free.

The main contributions of this paper are the following:
\begin{itemize}
    \item We propose a novel transmit waveform model to be used by distributed \ac{ISAC} \acp{AP} in a cell-free \ac{MIMO} network.
    \item We calculate the optimal transmit waveform by solving a constrained \ac{CRB} minimization problem. The relaxed version of the original problem is proved to be tight for the scenario at hand
    \item We characterize the optimal \ac{AN} structure, showing that it's rank-1 and  directed towards the eve.
    \item Numerical simulation prove the trade-off between sensing and communication performances and show the effects that the \acp{UE}-Eve proximity has on the optimal \ac{CRB}.
\end{itemize}

\section{System model}\label{system model}

Consider a cell-free \ac{MIMO} network  consisting of $M$  \acp{AP}, $K$  \acp{UE} and a sensing target, assumed to be an eve. To protect the privacy and security of the data traffic from the Eve, we introduce an \ac{AN} vector $\boldsymbol{\xi}$ \cite{su2022sensing} that will be used to reduce the sensing target's \ac{SNR}. 
Each \ac{AP} is equipped with a \ac{ULA} comprising $N$ transmit and $N$ receive antennas. 
The inter-element spacing within the \ac{ULA} is set to $\lambda/2$, where $\lambda$ is the carrier wavelength. Conversely, each  \ac{UE} is equipped with a single antenna. Digital beamforming capabilities are assumed for each AP, which is connected to a \ac{CPU} responsible for designing the transmit signals. In each timeslot $t$,  \ac{AP}~$m$ transmits a waveform denoted by the complex vector $\boldsymbol{\phi}_m^t \in \mathbb{C}^N$, which can be modeled  as
\begin{align}\label{small phi t}
\boldsymbol{\phi}_m^t=\sum_{s=1}^S\boldsymbol{\phi}_{m,s}^t = \sum_{s=1}^S\left(\ff_{m,s} x_{m,s}^t + \boldsymbol{\xi}_{m,s}^t\right),
\end{align}
%where $\FF_{m}=[\ff_{m,1}\dots\ff_{m,S}]\in \mathbb{C}^{N\times S}$ 
where $\ff_{m,s} \in \mathbb{C}^N$ is the precoding vector used by \ac{AP}~$m$ for stream $s$.
$S$ represents the number of transmission streams: each \ac{UE} is served by one transmission stream and an additional stream is allocated for sensing purposes. Consequently, $S=K+1$.
On the other hand, $x_{m,s}^t$ represents the complex-valued unit-power information symbol and the vector $\boldsymbol{\xi}_{m,s}^t \in \mathbb{C}^N$ represents the AN vector transmitted during timeslot $t$ by \ac{AP} $m$ on stream $s$. 
This noise is statistically independent of the transmitted symbols. In order to maximize its entropy, the noise vectors are modeled as complex Gaussian vectors.
We assume that  AN vectors belonging to different streams are uncorrelated with each other and that all $\boldsymbol{\xi}_{m,s}^t$ have the same covariance matrix, denoted by $\frac{\Rlam}{S}$.
We can get rid of the summation and express \eqref{small phi t} using matrices:
\vspace{-1mm}
 \begin{align}\label{small phi matrix}
       \boldsymbol{\phi}_m^t = \FF_{m} \xx_m^t+\boldsymbol{\xi}_m^t,
 \end{align}
 %\vspace{-3mm}
where $\boldsymbol{\xi}_m^t=\sum_{s=1}^S\boldsymbol{\xi}_{m,s}^t,~\FF_{m}=[\ff_{m,1}\dots\ff_{m,S}]\in \mathbb{C}^{N\times S}$ and $\xx_m^t = [x_{m,1}^t \ldots x_{m,S}^t ]^\top \in \mathbb{C}^{S}$, 
All the following derivations assume that symbols belonging to different streams and different \acp{AP} are statistically uncorrelated, i.e.,~$\mathbb{E}[(x_{m,s}^t)^\hermit (x_{m',s'}^t)] = 1$ if and only if $m=m'$ and $s=s'$.
This condition can be achieved through pseudo-random coding \cite{liu2020joint}.

\subsection{Communication observation}
The channel vector between  \ac{UE} $k$ and \ac{AP} $m$ is denoted by the vector $\hh_{m,k} \in \mathbb{C}^{N} $.
This allow us to define the received signal observed by \ac{UE} $k$ during the timeslot $t$ as 
\begin{align}\label{communication observation}
     y_{k}^{t,\text{c}} &= \sum_{m=1}^M \hh_{m,k}^\hermit \boldsymbol{\phi}_m^t+ w_u =\sum_{m=1}^M \hh_{m,k}^\hermit \ff_{m,k}x_{m,k}^t +\\
    &\sum_{\substack{s \neq k\\s=1}}^S \sum_{m=1}^M \hh_{m,k}^\hermit \ff_{m,s}x_{m,s}^t \nonumber 
    +\sum_{m=1}^M \hh_{m,k}^\hermit\boldsymbol{\xi}_m^t  + w_k,
\end{align}
where $w_k\sim \mathcal{CN}(0,\sigma_\text{c}^2)$ is the receiver noise, assumed to be uncorrelated from the transmitted waveform.

 \begin{figure}[t]
     \centering
     \input{images/general_scenario/Scenario_figure.tikz}
     \caption{illustration of the Cell-free MIMO network where ISAC APs serve communication users while sensing an Eve}
     \label{scenario}
\end{figure}
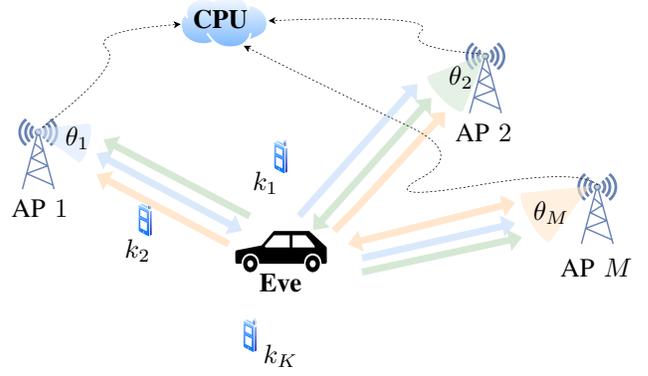

\subsection{Multistatic sensing}
As shown in Fig.~\ref{scenario}, the \acp{AP} implement multistatic sensing: every \ac{AP} receives the echo of its own transmitted signal plus the reflections coming from all of the other \acp{AP}.
The Eve is modeled as a point reflector having  a \ac{LoS} path with every \ac{AP}.
Under these assumptions, we can express the radar observations collected by \ac{AP} $m$ during timeslot $t$ as \cite{demirhan2023cell}
\begin{align}\label{radar observation}
  \yy_{m}^{t,\text{s}}&=\sum_{s=1}^S \underbrace{\sum_{m'=1}^M  \alpha_{m'}^m \atx(\theta_m)\atx(\theta_{m'})^\hermit  \boldsymbol{\phi}_{m',s}^t} _{\mathbf{g}_{m,s}^t} 
    + \mathbf{n}_m \nonumber\\
    &= \sum_{s=1}^S \sum_{m'=1}^M    \alpha^m_{m'} \atx(\theta_m)\atx(\theta_{m'})^\hermit \ff_{m',s} x_{m',s}^t \nonumber \\
    &+\sum_{m'=1}^M\alpha_{m'}^m \atx(\theta_m)\atx(\theta_{m'})^\hermit\boldsymbol{\xi}_{m'}^t + \mathbf{n}_m,
\end{align}
%%% 
where $|\alpha_{m'}^m|^2$ is the channel gain between the transmitting \ac{AP} $m'$ and receiving \ac{AP} $m$. To take into account the effects of pathloss and \ac{RCS}, the channel gain is modeled according to the Swerling-I model, that is  $\alpha_{m'}^m \sim \mathcal{CN}(0,(\delta_{m'}^m)^2)$.
The angles $\theta_m$ and $\theta_{m'}$ represent the \ac{AoA} to \ac{AP} $m$ and the \ac{AoD} from \ac{AP} $m'$, whereas $\atx(\cdot)$ is the steering vector of a ULA.
The thermal noise vector is denoted by $\mathbf{n}_m \sim \mathcal{CN}(0,\sigma_\text{s}^2 \mathbf{I}_N)$, and is assumed to be uncorrelated from the transmitted waveform.

\section{Transmitted waveform joint design}
The sensing objective we have chosen to minimize is the \ac{CRB} on $\theta_1, \ldots \theta_m$.
Assuming that the optimal waveform is computed during timeslot $t$, we assume that $\theta_1, \ldots \theta_m$ have been estimated (e.g. using the MUSIC algorithm with omnidirectional pilots \cite{zahernia2011music}) 
during timeslot $t-1$ and that, given a low mobility of the Eve, said estimations still hold true during the present timeslot. 
In this way, using the initial estimation, The cell-free \ac{MIMO} network can compute a transmit waveform that boosts its future estimation performances while simultaneously degrading the Eve channel quality by means of the \ac{AN}.
In this section, we first derive the \ac{FIM} and \ac{CRB} and then use them to define the optimization problem trough which we design the optimal transmitted waveform, \ac{SDR} is used to obtain a relaxed problem solvable trough complex optimization solvers.
\subsection{\ac{CRB} derivation} 
Let $\boldsymbol{\eta}\in  \mathbb{R}^{(2M^2 +M)}$ denote the vector of unknown channel parameters 
\vspace{-1mm}
\begin{align}
    \boldsymbol{\eta}=[\Re\{\alpha_{1}^1\},\Im\{\alpha_{1}^1\},\ldots,\Re\{\alpha_{M}^M\},\Im\{\alpha_{M}^M\},
     \theta_1,\ldots,\theta_M]^\top ,
\end{align}
\vspace{-2mm}

where $\Re\{\alpha_{1}^1\}$ and $\Im\{\alpha_{1}^1\}$ represent the real and imaginary part of the complex channel coefficient $\alpha_{1}^1$
Let us then define the vectors $\yy^{t,\text{s}} =[(\yy_{1}^{t,\text{s}})^\top \ldots (\yy_{M}^{t,\text{s}})^\top ]^\top \in \mathbb{C}^{MN}$ and $\mathbf{g}^t_s =[(\mathbf{g}^t_{1,s})^\top\ldots (\mathbf{g}^t_{M,s})^\top ]^\top \in \mathbb{C}^{MN}$.  The \ac{FIM} $  \boldsymbol{J_\eta} \in \mathbb{R}^{(2M^2 +M) \times (2M^2 +M)}$ can be then calculated as %\cite{kay1993fundamentals}
\begin{align}
    \boldsymbol{J_\eta} &= \mathbb{E}_{\yy^{t,\text{s}}| \boldsymbol{\eta}}\Bigg[ -\frac{\partial^2 \log f(\yy^{t,\text{s}}| \boldsymbol{\eta})}{\partial \boldsymbol{\eta}\partial\boldsymbol{\eta}^\hermit} \Bigg]\nonumber\\
    &= \frac{2}{\sigma_\text{s}^2}  \sum_{s=1}^{S} \Re \Bigg\{  \bigg( \frac{\partial \mathbf{g}^t_s}{\partial  \boldsymbol{\eta}} \Bigg)^\hermit \bigg( \frac{\partial \mathbf{g}^t_s}{\partial  \boldsymbol{\eta}} \Bigg)                     
    \Bigg\},
\end{align}
where $f(\yy^{t,\text{s}}| \boldsymbol{\eta})$ is the conditional likelihood function of the stacked observation vector $\yy^{t,\text{s}}$ given $\boldsymbol{\eta}$.
The \ac{CRB} for the estimation of $\theta_m$ is then defined as
\begin{align}
\textrm{CRB}_{\theta_m}=\sqrt{[\boldsymbol{J_\eta}^{-1}]_{2M^2+m,2M^2+m}}.
\end{align}
The square root has been included to facilitate future comparison with the estimation root mean square error.

\subsection{Optimal transmit waveform}
The chosen optimization metric is practically translated into the minimization of the \ac{FIM}'s inverse trace.
The optimization variables are the \ac{AP} precoding matrices $\FF_m$ and the \ac{AN} covariance matrices $\Rlam$.
In order to define the constraints of the optimization problem, we first need to mathematically define the \ac{UE} \ac{SINR} and the Eve \ac{SNR}.
The communication channel $\hh_{m,k}$ is assumed to be deterministic, a condition achievable through the acquisition of channel-state information at the transmitter, and the multi-user interference is treated as noise \cite{demirhan2023cell}.
Following \eqref{communication observation}, the \ac{SINR} for user $k$ can be defined as 
\begin{align}\label{SINR}
        &\text{SINR}_k =\\
        &\frac{ \sum_{m=1}^M |\mathbf{h}_{m,k}^\text{H} \mathbf{f}_{m,k} |^2}
    {\sum_{m=1}^M \sum_{\substack{s \neq k\\s=1}}^S |\mathbf{h}_{m,k}^\text{H} \mathbf{f}_{m,s} |^2 +  
    \sum_{m=1}^M |\mathbf{h}_{m,k}^\text{H}\Rlam \mathbf{h}_{m,k} |+\sigma_\text{c}^2  } \nonumber .
\end{align}
On the other hand, the eve's \ac{SNR}  can be modelled as 
\begin{align}\label{SNR}
        \text{SNR}_{\text{E}}=\frac{\sum_{m=1}^M \sum_{s=1}^S  (\delta_{m}^m)^2|\atx(\theta_m)^\text{H}\ff_{s,m}|^2}{\sum_{m=1}^M  (\delta_{m}^m)^2|\atx(\theta_m)^\text{H}\Rlam\atx(\theta_m)|+\sigma_\text{s}^2}.
\end{align}
We are now able to define the proposed transmit waveform design strategy
 \begin{subequations}\label{original problem}
 \begin{align}
     \underset{\{\Rlam\}_1^M , \{\ff_{m,s}\}_{1,1}^{M,S}}{\mathrm{minimize}}&\text{Tr}(\mathbf{J}_{\boldsymbol{\eta}}^{-1}),\\ 
     \text{subject to} \quad~~& \text{SINR}_k\geq \gamma_k,~ k= 1, \ldots ,K \label{SINR_constr_og}\\
     &\text{SNR}_{\text{E}} \leq \psi, \label{SNR_constr_og} \\
     &|| \boldsymbol{\phi}_t || ^2 \leq P_m  ,~ m=1 \ldots M ,\label{power_constr_og}
 \end{align}
 \end{subequations}
 where $P_m, \gamma_k$ and $\psi$ represent the available power at \ac{AP} $m$, the required \ac{SINR} for user $k$, and the maximum \ac{SNR} that the Eve should be allowed to achieve.

 \subsection{Semidefinite reformulation}
 Problem \eqref{original problem} is non-convex in the variable $\ff_{m,s}$ due to the quadratic terms occurring in both the \ac{SINR} and \ac{SNR}.
To solve this, we reformulate \eqref{original problem} as an \ac{SDP}\cite{gershman2010convex}. In particular, we define the variable $\WW_s=\ff_s\ff_s^{\text{H}}$, where $\ff_s=[\ff_{1,s}^\top \dots \ff_{M,s}^\top]^\top \in \mathbb{C}^{ NM }$ and express the objective function and the constraints as a function of this variable.
We can then reformulate
\eqref{SINR_constr_og} as 
\begin{align}\label{SINR_sdp}
    &\text{Tr}(\hh_k\hh_k^\text{H}\WW_k) \geq \\
    & \gamma_k \Big(\sum_{\substack{s \neq k\\s=1}}^S \text{Tr}(\hh_k\hh_k^\text{H} \WW_{s})  +  
    \sum_{m=1}^M \text{Tr}(\hh_{m,k}\hh_{m,k}^\text{H} \Rlam) \ +              
   \sigma^2_\text{c}\Big) ,\nonumber
\end{align}
where $\mathbf{h}_k=[\mathbf{h}_{1,k}^\top \dots \mathbf{h}_{M,k}^\top ]^\top \in \mathbb{C}^{NM}$.
Similarly, the Eve's \ac{SNR} constraint in \eqref{SNR_constr_og} can be redefined as 

\begin{align}\label{SNR_sdp}
    &\sum_{m=1}^M \sum_{s=1}^S (\delta_{m}^m)^2 \text{Tr}(\atx(\theta_m)\atx(\theta_m)^\text{H}\WW_{m,s} )    \leq \nonumber\\
     &\psi\Big(\sum_{m=1}^M  (\delta_{m}^m)^2\text{Tr}(\atx(\theta_m)\atx(\theta_m)^\text{H}\Rlam )   +\sigma_\text{s}^2\Big),
\end{align}
where $\WW_{m,s}=\ff_{m,s}\ff_{m,s}^\hermit$ and this is the $m^{\text{th}}$ $N\times N$ block onto the main diagonal of $\WW_s$.
Leveraging the lack of correlation between the terms in \eqref{small phi matrix}, \eqref{power_constr_og} can be re-written  as:
\begin{align}\label{power_constr_sdp}
||\boldsymbol{\phi}_m^t||^2 &=%\text{Tr}(\boldsymbol{\phi}_m^t  (\boldsymbol{\phi}_m^t)^\hermit ) 
 %= \text{Tr} (\FF_m \FF_m^\hermit + \Rlam) \nonumber\\
\text{Tr} \Bigg(\Bigg(\sum_{s=1}^S \WW_{m,s}\Bigg) + \Rlam\Bigg) \leq P_m. 
\end{align}
Eventually, the \ac{SDP}-equivalent problem can be defined as : 
 \begin{subequations}\label{sdp problem}
 \begin{align}
    \underset{\{\Rlam\}_1^M , \{\WW_s\}_{1}^{S}}{\mathrm{minimize}}&\text{Tr}\Big(\mathbf{J}_{\boldsymbol{\eta}}^{-1}\Big)\\ 
     \text{subject to} \quad~& \eqref{SINR_sdp}~ k= 1 \ldots 
     K, \\
     &\eqref{SNR_sdp}  \\
     &\eqref{power_constr_sdp}~ m=1 \ldots M,  \\
     & \text{rank}(\WW_s)=1,~~s=1\ldots S \label{rank-1 constraint} \\
     & \WW_s \in \mathbb{S}^+,~~s=1\ldots S, 
 \end{align}
 \end{subequations}
 where $\mathbb{S}^+$ represents the set of Hermitian positive semidefinite matrices, the reformulation of $\mathbf{J}_{\boldsymbol{\eta}}$ elements as a function of the relaxed variables is shown in the appendix.
 This problem can be relaxed by removing constraint \eqref{rank-1 constraint} and then solved via convex optimization solvers \cite{luo2010semidefinite}: 
  \begin{subequations}\label{sdp-realxed problem}
 \begin{align}
      \underset{\{\Rlam\}_1^M , \{\WW_s\}_{1}^{S}}{\mathrm{minimize}}&\text{Tr}\Big(\mathbf{J}_{\boldsymbol{\eta}}^{-1}\Big)\\ 
     \text{subject to} \quad& \eqref{SINR_sdp}~ k= 1 \ldots 
     K,  \\
     &\eqref{SNR_sdp} \\
     &\eqref{power_constr_sdp}~ m=1 \ldots M,  \\
     & \WW_s \in \mathbb{S}^+,~~s=1\ldots S. 
 \end{align}
   \end{subequations}
In the next section we show that the optimal solution of problem \ref{sdp-realxed problem} yields rank-$1$ matrices, making said solution optimal for problem \ref{sdp problem} and the relaxation tight.

\section{Simulation results}

\begin{figure}[t!]
     \centering
     \input{images/AP_scheme/AP_scheme.tikz}
     \caption{simulated APs and UEs layout, the UEs are randomly placed withing the green box}
     \centering
     \label{AP scheme}
\end{figure}
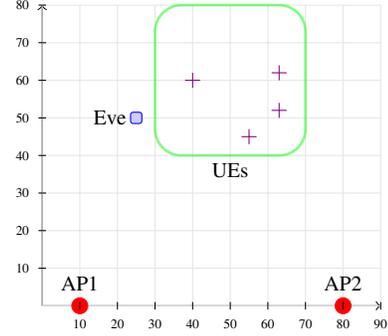
 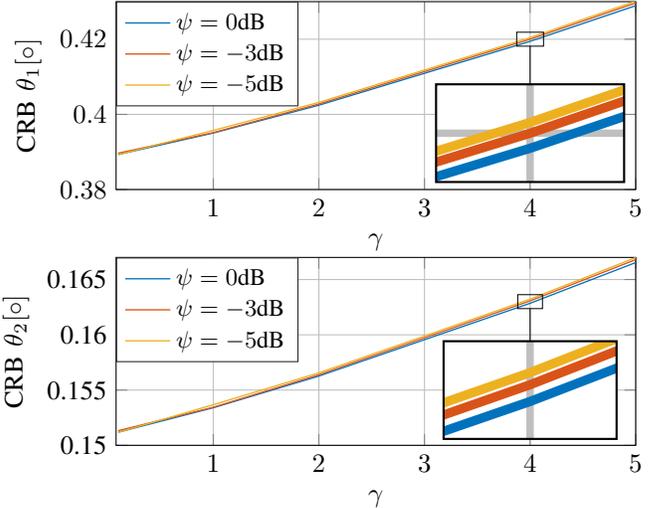
\begin{figure}[t!]
     \centering
     \input{images/CRB_vs_psi/CRB_vs_psi}
     \vspace{-2mm}
     \caption{\ac{AP} $1$ (top) and \ac{AP} $2$ (bottom) \ac{CRB} on $\theta_m$ for different $\psi$}
     \centering
     \vspace{-2mm}
     \label{CRB vs psi}
\end{figure}
\vspace{5mm}
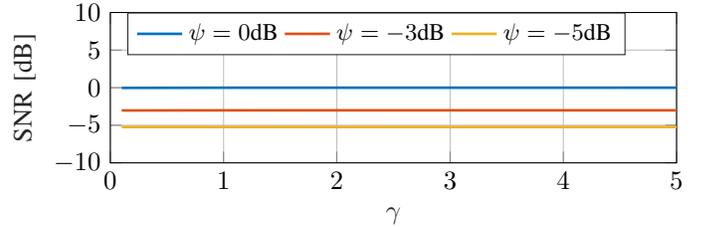
\begin{figure}[t!]
     \centering
     \input{images/SNR_vs_psi/SNR_vs_psi}
     \vspace{-5mm}
     \caption{achieved eve's SNR vs $\gamma$ for different values of $\psi$}
     \centering
     \label{SNR vs psi}
\vspace{-6mm}     
\end{figure}

In this section, we now present the performance of the proposed waveform design strategy in a cell-free \ac{MIMO} network with $M=2$ \acp{AP}, each of them equipped with a \ac{ULA} of $N=30$ antennas, using the same waveform to  serve $K=4$ \acp{UE} and sense one Eve.
The \acp{UE} are randomly distributed in a $40 \times 40 $\,m area in front of the \acp{AP}, located at the coordinates $[10,0]$ and $[80,0]$.
We then choose $P_m=1$\,W the noise variances are $\sigma^2_\text{c} =\sigma^2_\text{s}=1 $ .
The Swerling-I model variance $(\delta_{m'}^m)^2$ has been set to $0.1$ for all the $m, m'$ pairs. Unless otherwise specified, the maximum 's \ac{SNR} $\psi$ has been set to $0$\,dB.
All the \acp{UE} are envisioned to be guaranteed the same \ac{SINR}, thus $\gamma_1=\ldots=\gamma_K=\gamma$

\subsection{CRB dependency by $\psi$}
In this subsection we investigate how the choice of  $\psi$  affects the achieved \ac{CRB} on $\theta_1$ and $\theta_2$.
Fig. \ref{CRB vs psi} shows the achieved \ac{CRB} as a function of $\gamma$ for three different $\psi$: We observe that larger $\gamma$ values correspond to higher \ac{CRB} values.
This is a manifestation of the sensing-communication trade-off: as the \ac{SINR} requirements become more demanding, the system needs to allocate more resources on the communication side, thus degrading the sensing performances, namely achieving a higher CRB. A second trade-off arises between the CRB and $\psi$: a more stringent constraint onto the Eve's \ac{SNR} corresponds to worse sensing performances. 
This can be ascribed to the direct proportionality between the system's sensing capabilities, namely the CRB, and the amount of power beamformed towards the target, i.e. the SNR.
Fig. \ref{CRB vs psi} shows that constraining the sensing SNR to $\psi=-5$ dB or $\psi=-3$ cause a proportional albeit very small CRB increase.
%higher CRB than the case where $\psi=0$ dB, whereas setting  dB generates a CRB very similar to the former.
This implies that a $3$ dB or even a $5$dB decrease of the allowed maximum sensing SNR is causing little to no degradation in the sensing performance, allowing us to set more stringent Eve's SNR contraints at almost no price.
This might instill the suspicion that constraint \eqref{SNR_sdp} is not tight, however 
Fig. \ref{SNR vs psi} indicates the contrary as the achieved Eve's \ac{SNR} is equal to $\psi$.

%Therefore,SNR to be lower than $\psi$ we are decreasing the system's sensing capabilities, thus generating a higher lower bound on the estimation variance.

\subsection{Relaxation's tightness}
The optimal  $\WW_s$ for the communication streams (i.e. $s=1 \ldots K$) are all rank-$1$ matrices.The communication beamforming vectors can be retrieved as  
\begin{align}\label{eigenvalue decomp} 
    &\ff_{s,m}=[\sqrt{\epsilon_s}\boldsymbol{u}]_{(m-1)N+1~:~mN},\\
    &m=1\ldots M,~s=1\ldots K, \nonumber
\end{align}
where $\epsilon_k$ and $\boldsymbol{u}$ are the non-zero eigenvalue and respective eigenvector associated to $\WW_s$.
As for the sensing stream ( i.e. $s=K+1$ ) The optimal $\WW_s$ is not rank-1 as  all its eigenvalues are in the order of $10^{-5}$ : We can then say that $\WW_s$ has rank-almost zero, making the presence of a sensing stream superfluous.
Lastly, the The optimal \ac{AN} covariance matrices $\{\Rlam^\text{opt} \}_1^M$ are rank-1: this result is to be expected as the \ac{AN} needs to be concentrated solely towards the Eve.
Given the previous considerations we can claim that a transmit waveform originated by the solution of problem \eqref{sdp-realxed problem} constitutes an optimal of problem \eqref{sdp problem}, thus making the relaxation tight.

\subsection{Artificial noise characterization}
\begin{figure}[t!]
     \centering
     \input{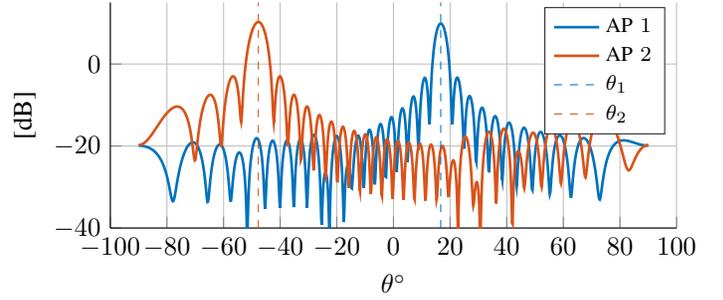}
     \vspace{-6mm}
     \caption{optimal \ac{AN} beampattern obtained for $\gamma=2$}
     \label{AN_bp_gamma4}
 \end{figure}

It can be seen that the \ac{AN} covariance matrix $\Rlam$ is used both in the \ac{UE} \ac{SINR} and in the Eve \ac{SNR} definitions.
In both equations, $\Rlam$ appears in the denominator, either multiplied by the communication channel vector $\mathbf{h}_{m,k}$ or the steering vector $\atx(\theta_m)$. The latter multiplication inevitably gives $\Rlam$ a directional selectivity, as shown in Fig. \ref{AN_bp_gamma4}

 \begin{figure}[t!]
     \centering
     \input{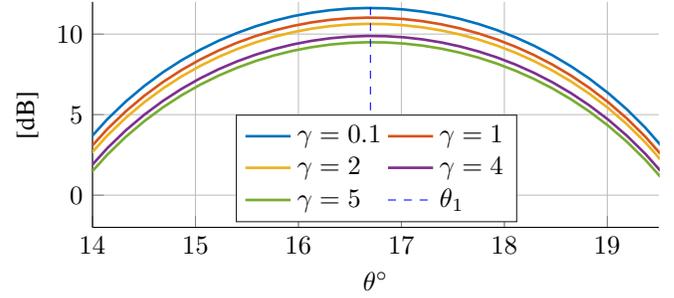}
     \vspace{-2mm}
     \caption{\ac{AP} $1$'s optimal \ac{AN} beampattern for different values of $\gamma$ }
     \label{AN_bp vs gamma}
 \end{figure}
 
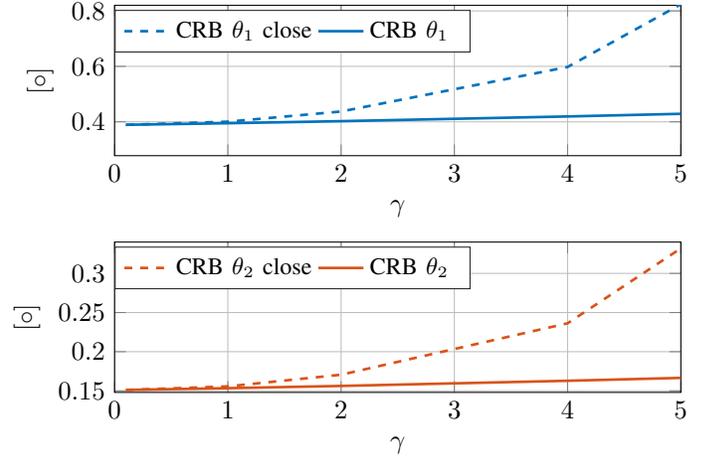
\begin{figure}[t!]
    \centering
    \input{images/close_vs_distant/CloseVSDistantSubplot}
    \vspace{-2mm}
    \caption{\ac{AP} $1$ (top) and \ac{AP} $2$ (bottom) \ac{CRB} on $\theta_m$ vs $\gamma$ for different \acp{UE}-Eve proximity conditions}
    \label{close_vs_distant}
    \vspace{-4mm}
\end{figure}

The \ac{AN} beam pattern, calculated as $|\atx(\theta)^H \Rlam|^2 $, peaks around $\theta_m$ with a rather narrow beamwidth and is otherwise below $-15$dB,in line with the fact that $\Rlam$ is rank-1.
Another manifestation of the previously mentioned performance tradeoff can be seen in Fig. \ref{AN_bp vs gamma}, where it's shown that an increase in $\gamma$ corresponds to a decrease in the $\ac{AN}$ peak value.
A possible explanation for this is that as the system tries to guarantee a higher \ac{SINR} to its users, the term $|\mathbf{h}_{m,k}^\text{H}\Rlam \mathbf{h}_{m,k} |$ must be lower, leading to a progressive decrease of the beampattern peak.

\subsection{\ac{CRB} vs \ac{SINR} and  proximity}
 Fig. \ref{close_vs_distant}, two curves are shown, labeled "distant" and  "close", respectively. (The "distant" configuration is shown in Fig. \ref{AP scheme}.)
The "close" configuration was created by placing two of the \acp{UE} at a distance of $0.5$\,m from the target, one on the left and one on the right. This configuration has been considered to investigate how the optimization problem would change if one or more \acp{UE} angular coordinates would fall within the \ac{AN} main lobe depicted in Fig. \ref{AN_bp_gamma4}.
Fig. \ref{close_vs_distant} confirms the logical intuition that such a proximity between the Eve and one or more \ac{UE} would inevitably lead to a degradation of the optimization function value.
It is also interesting to note that the \ac{CRB}'s proximity-induced degradation  is proportional to $\gamma$. This behavior can be intuitively attributed to the fact that the system can allocate more resources to the Eve when $\gamma$ is small, effectively mitigating the disadvantage of proximity.
However, when $\gamma$ becomes larger, the \ac{SINR} constraint becomes tighter and the system can no longer allocate additional resources to compensate for the proximity, resulting in a higher \ac{CRB}.

\section{Conclusions}
In this paper, we have considered a secure transmit waveform design used by \ac{ISAC} \acp{AP} in a cell-free \ac{MIMO} environment to serve the communication users while preventing an Eve from spoofing the communication information. 
To do this, the transmitted dual waveform includes an \ac{AN} vector whose goal is to degrade the Eve's \ac{SNR}.
The optimal precoding and \ac{AN} covariance matrices  are computed by minimizing the \ac{CRB} on the monostatic observation angles under SINR constraints for the communication users and maximum SNR constraint for the Eve.
Our simulations have shown the inherent tradeoff between sensing and communication performances and how, given the angular directionality of the \ac{AN},\acp{UE}-Eve proximity degrade the system's performances, making this design strategy very efficient in limiting the impact of an Eve that's physically separated from the legitimate communication users. 

\appendix
\vspace{-2mm}
\section{\ac{FIM} as a function of $\WW_s$}
Here we show how the \ac{FIM} elements can be expressed as a function of the \ac{SDP} variable $\WW_s$:
\begin{align}
    &\Bigg(\frac{\partial  \mathbf{g}^t_s }{  \partial \Re\{\alpha_m^{m}\}}\Bigg)^\hermit  \frac{\partial  \mathbf{g}^t_s }{  \partial \Re\{\alpha_m^{m}\}} = \\&(\boldsymbol{\phi}_{m,s}^t )^\hermit \atx(\theta_{m})\atx(\theta_m)^\hermit \atx(\theta_m)\atx(\theta_{m})^\hermit \boldsymbol{\phi}_{m,s}^t= \nonumber\\
 &N\textrm{Tr}\Big(\Big(\WW_{m,s}+\frac{\Rlam}{S}\Big)\atx(\theta_{m})\atx(\theta_{m})^\hermit\Big),\nonumber
\end{align}
where  the last form leverages  the uncorrelation between the components of $\boldsymbol{\phi}_{m,s}^t$, as shown in section \ref{system model}.
Let us now move on and present the second type of product
 \begin{align}
     &\Bigg(\frac{\partial  \mathbf{g}^t_s }{  \partial \Re\{\alpha_m^{m}\}}\Bigg)^\hermit  \frac{\partial  \mathbf{g}^t_s }{  \partial \theta_m}=\sum_{\Tilde{m}=1}^M \Bigg(\frac{\partial  \mathbf{g}^t_{\Tilde{m},s} }{  \partial \Re\{\alpha_m^{m}\}}\Bigg)^\hermit  \frac{\partial  \mathbf{g}^t_{\Tilde{m},s} }{  \partial \theta_m} = \nonumber\\
     &\Bigg(\frac{\partial  \mathbf{g}^t_{m,s} }{  \partial \Re\{\alpha_m^{m}\}}\Bigg)^\hermit  \frac{\partial  \mathbf{g}^t_{m,s} }{  \partial \theta_m}=(\boldsymbol{\phi}_{m,s}^t )^\hermit \atx(\theta_{m})\atx(\theta_m)^\hermit \\
     & \times\Bigg(\alpha_m^m \Big(\atxdt(\theta_m)\atx(\theta_m)^\hermit +\atx(\theta_m)\atxdt(\theta_m)^\hermit\Big)\boldsymbol{\phi}_{m,s}^t + \nonumber\\ 
       &\sum_{\substack{m' \neq m\\m'=1}}^M \alpha_{m'}^{m} \atxdt(\theta_m)\atx(\theta_{m'})^\hermit \boldsymbol{\phi}_{m',s}^t   \Bigg) =\alpha_m^{m}\textrm{Tr}\Bigg(\Big(\WW_{m,s}+\frac{\Rlam}{S}\Big)\nonumber\\
& \times\atx(\theta_{m})\atx(\theta_m)^\hermit\Big(\atxdt(\theta_m)\atx(\theta_m)^\hermit +\atx(\theta_m)\atxdt(\theta_m)^\hermit\Big)\Bigg),\nonumber
 \end{align}
 where we use the fact that $\mathbb{E}[(x_{m,s}^t)^\hermit (x_{m',s'}^t)] = 0$.
 The notation $\atxdt(\cdot)$ denotes the steering vector's first derivative, defined as $[\atxdt(\theta_m)]_n=j\pi n \cos(\theta_m) e^{j\pi n \sin(\theta_m)}$.
 Moving on, the next product is defined as 
\begin{align}
    &\Bigg(\frac{\partial  \mathbf{g}^t_{s} }{  \partial \theta_{m}}\Bigg)^\hermit  \frac{\partial  \mathbf{g}^t_{s} }{ \partial \theta_m} = \Bigg(\frac{\partial  \mathbf{g}^t_{m,s} }{  \partial \theta_{m}}\Bigg)^\hermit   \frac{\partial  \mathbf{g}^t_{m,s} }{ \partial \theta_m} +
    \sum_{\substack{\Tilde{m}=1 \nonumber\\ \Tilde{m} \neq m}}^M \Bigg(\frac{\partial  \mathbf{g}^t_{\Tilde{m},s} }{  \partial \theta_{m}}\Bigg)^\hermit  \frac{\partial  \mathbf{g}^t_{\Tilde{m},s} }{ \partial \theta_m}  \\
    &=|\alpha_m^m|^2 \textrm{Tr}\Bigg(\Big(\WW_{m,s}+\frac{\Rlam}{S}\Big)\Big(\atxdt(\theta_m)\atx(\theta_m)^\hermit +\atx(\theta_m)\atxdt(\theta_m)^\hermit\Big)^\hermit\nonumber\\ &\times\Big(\atxdt(\theta_m)\atx(\theta_m)^\hermit +\atx(\theta_m)\atxdt(\theta_m)^\hermit\Big) \Bigg) +  \nonumber \\
    &\sum_{\substack{m'=1 \\ m'\neq m}}^M|\alpha^{m}_{m'}|^2 \textrm{Tr}\Bigg(\Big(\WW_{m',s}+\frac{\Rlamprime}{S}\Big)\atx(\theta_{m'})\atxdt(\theta_m)^\hermit \atxdt(\theta_m)\atx(\theta_{m'})^\hermit \Bigg)\nonumber + \\
    &\sum_{\substack{m'=1 \\ m'\neq m}}^M N |\alpha_m^{m'}|^2 \textrm{Tr}\Bigg(\Big(\WW_{m,s}+\frac{\Rlam}{S}\Big)\atxdt(\theta_m)^\hermit \atxdt(\theta_m) \Bigg).
\end{align} 

% where
% \begin{align}
%     \frac{\partial  \mathbf{g}^t_{\Tilde{m},s} }{  \partial \theta_{m}} =  \alpha_{\Tilde{m}}^m \atx(\theta_{\Tilde{m} })\atxdt(\theta_m)^\hermit\boldsymbol{\phi}_{m,s}^t
% \end{align}
All the other kind of products appearing in the \ac{FIM}
have been omitted due to space constraints.

\bibliographystyle{IEEEtran}
\bibliography{auxiliary/biblio}

\end{document}

%% file: auxiliary/commands.tex
\newcommand{\yy}{\mathbf{y}}
\newcommand{\xx}{\mathbf{x}}

\newcommand{\WW}{\mathbf{W}}

\newcommand{\ff}{\mathbf{f}}
\newcommand{\hh}{\mathbf{h}}
\newcommand{\FF}{\mathbf{F}}

\newcommand{\Rlam}{\mathbf{R}_{\mathbf{\Xi}_m} }
\newcommand{\Rlamprime}{\mathbf{R}_{\mathbf{\Xi}_{m'}} }

\newcommand{\hermit}{\mathsf{H}}

\usepackage{accents}
\newcommand*{\dt}[1]{%
	\accentset{\mbox{\large .}}{#1}}

\newcommand{\atxdt}{\dt{\mathbf{a}} }

\newcommand{\atx}{\mathbf{a}}

%% file: auxiliary/acronyms.tex
\acrodef{MIMO}[MIMO]{multiple-input multiple-output}
\acrodef{AP}[AP]{access point}
\acrodef{UE}[UE]{user equipment}
\acrodef{ULA}[ULA]{uniform linear array}
\acrodef{CPU}[CPU]{central processing unit}
\acrodef{LoS}[LoS]{line of sight}
\acrodef{RCS}[RCS]{radar cross section}
\acrodef{AoD}[AoD]{angle of departure}
\acrodef{AoA}[AoA]{angle of arrival}
\acrodef{CRB}[CRB]{Cramer-Rao bound}
\acrodef{FIM}[FIM]{Fisher information matrix}
\acrodef{AN}[AN]{artificial noise}
\acrodef{SINR}[SINR]{signal to noise and interference ratio}
\acrodef{SNR}[SNR]{signal to noise ratio}
\acrodef{QoS}[QoS]{quality of service}
\acrodef{SDR}[SDR]{semi-definite relaxation}
\acrodef{SDP}[SDP]{semi-definite program}
\acrodef{ISAC}[ISAC]{integrated sensing and communications}
\acrodef{PLS}[PLS]{physical layer security}
\acrodef{CSI}[CSI]{channel-state information}
\acrodef{MUI}[MUI]{multi-user interference}

%% file: images/general_scenario/Scenario_figure.tikz
\begin{tikzpicture}

    \node at (-2, 0) {\includegraphics[width=0.9\linewidth]{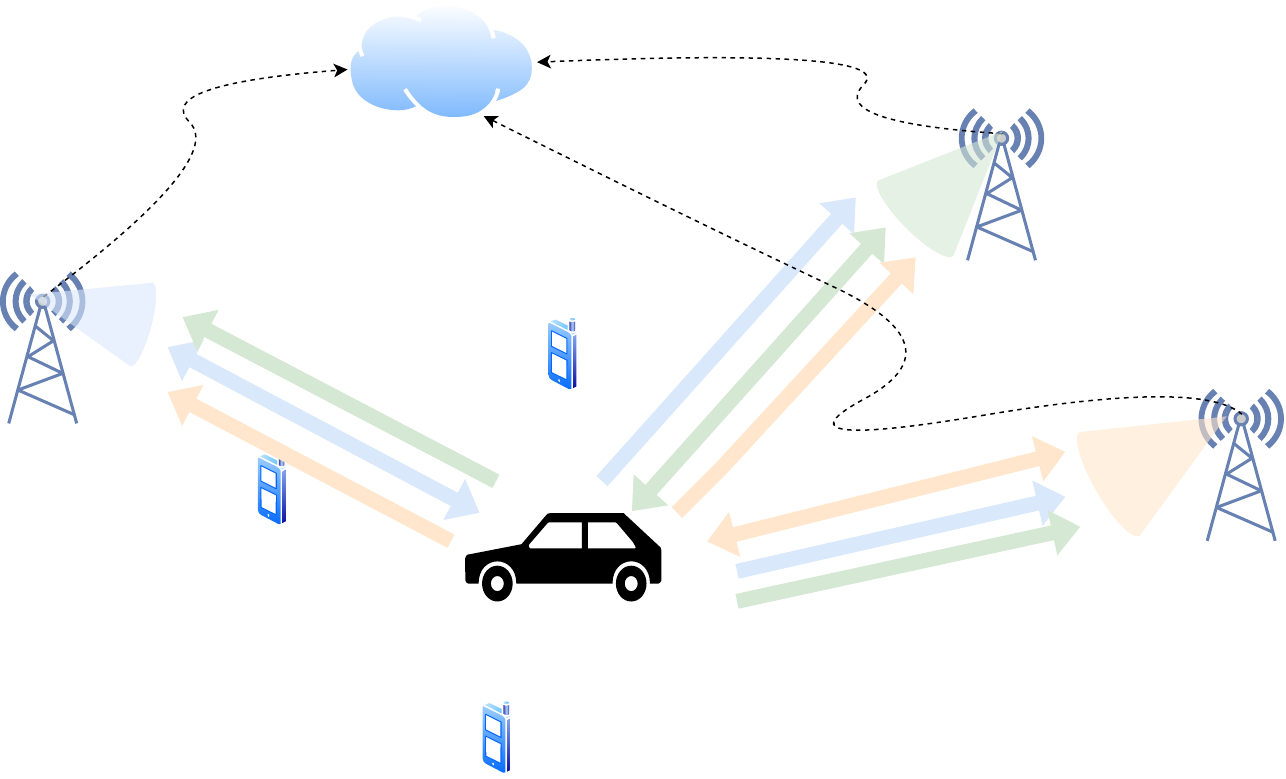}};
     \node[above] at (-3.3, 1.8) {\textbf{CPU}};
     \node[above] at (-5.7, -0.7) {AP $1$};
     \node[above] at (-5.2, 0.2) {$\theta_1$};
     \node[above] at (1.7, -1.5) {AP $M$};
     \node[above] at (1.1, -0.8) {$\theta_M$};
     \node[above] at (-0.1, 1) {$\theta_2$};
     \node[above] at (0.2, 0.3) {AP $2$};
     \node[above] at (-2.5, -1.7) {\textbf{Eve}};
     \node[above] at (-2.7, -0.4) {$k_1$};
     \node[above] at (-2.5, -2.7) {$k_K$};
     \node[above] at (-4.4, -1.3) {$k_2$};
    
    % Add your TikZ commands here
\end{tikzpicture}

%% file: images/AP_scheme/AP_scheme.tikz
\begin{tikzpicture}[scale=0.5]

% Define x and y limits
\draw[->] (0,0) -- (9,0) node[right] {};
\draw[->] (0,0) -- (0,8) node[above] {};
\draw[gray!20,step=1] (0,0) grid (9,8);

% Plot red dots
\draw[red, fill=red] (1,0) circle (6pt);
\draw[red, fill=red] (8,0) circle (6pt);
\draw[blue, fill=blue!20, rounded corners=1pt] (2.5-0.15,5-0.15) rectangle (2.5+0.15,5+0.15);

% Plot green box with rounded corners and adjusted opacity
\draw[green, line width=1pt, rounded corners=10pt,opacity=0.5] (3,4) rectangle (7,8);

% Draw x and y scales
\foreach \x in {1,2,...,9}
    \draw (\x,0.1) -- (\x,-0.1) node[below,font=\tiny] {\x0};
\foreach \y in {1,2,...,8}
    \draw (0.1,\y) -- (-0.1,\y) node[left,font=\tiny] {\y0};

% Add text "AP" in correspondence to the red dots
\node[font=\footnotesize] at (1,0.6) {AP1};
\node[font=\footnotesize] at (8,0.6) {AP2};
\node[align=center,font=\footnotesize] at (5,3.6) {UEs};
\node[align=center,font=\footnotesize] at (1.8,5) {Eve};

% draw the users

\pgfmathsetmacro{\rx}{4}
\pgfmathsetmacro{\ry}{6}
\draw[violet] (\rx-0.2, \ry) -- (\rx+0.2, \ry);
\draw[violet] (\rx, \ry-0.2) -- (\rx, \ry+0.2);

\pgfmathsetmacro{\rx}{5.5}
\pgfmathsetmacro{\ry}{4.5}
\draw[violet] (\rx-0.2, \ry) -- (\rx+0.2, \ry);
\draw[violet] (\rx, \ry-0.2) -- (\rx, \ry+0.2);

\pgfmathsetmacro{\rx}{6.3}
\pgfmathsetmacro{\ry}{6.2}
\draw[violet] (\rx-0.2, \ry) -- (\rx+0.2, \ry);
\draw[violet] (\rx, \ry-0.2) -- (\rx, \ry+0.2);

\pgfmathsetmacro{\rx}{6.3}
\pgfmathsetmacro{\ry}{5.2}
\draw[violet] (\rx-0.2, \ry) -- (\rx+0.2, \ry);
\draw[violet] (\rx, \ry-0.2) -- (\rx, \ry+0.2);

\end{tikzpicture}

%% file: images/CRB_vs_psi/CRB_vs_psi.tex
\definecolor{mycolor1}{rgb}{0.00000,0.44700,0.74100}%
\definecolor{mycolor2}{rgb}{0.85000,0.32500,0.09800}%
\definecolor{mycolor3}{rgb}{0.92900,0.69400,0.12500}%
\begin{tikzpicture}[spy using outlines={rectangle, magnification=7,connect spies}]

\begin{axis}[%
width=0.78\linewidth,
height=2.5cm,
at={(0.58in,2.54in)},
scale only axis,
xmin=0.08,
xmax=5,
xlabel style={font=\color{white!15!black}},
xlabel={$\gamma$},
ymin=0.38,
ymax=0.43,
ylabel style={font=\color{white!15!black}},
ylabel={CRB $ \theta_1 [\circ]$},
axis background/.style={fill=white},
xmajorgrids,
ymajorgrids,
legend columns=1,
legend style={at={(0,1)}, anchor=north west, legend cell align=left, align=left, draw=white!15!black,font=\small}
]
\addplot [color=mycolor1, line width=0.5pt]
  table[row sep=crcr]{%
0.1	0.389312366886668\\
1	0.395034965578866\\
2	0.402433237970152\\
4	0.41942681677198\\
5	0.42892500943948\\
};
\addlegendentry{$\psi=0$dB}

\addplot [color=mycolor2, line width=0.5pt]
  table[row sep=crcr]{%
0.1	0.389674003638887\\
1	0.395139482267171\\
2	0.40275857962847\\
4	0.419997896737746\\
5	0.429634342086065\\
};
\addlegendentry{$\psi=-3$dB}

\addplot [color=mycolor3, line width=0.5pt]
  table[row sep=crcr]{%
0.1	0.389272860248282\\
1	0.3957348089108\\
2	0.403188762736466\\
4	0.420409604719931\\
5	0.430120731605095\\
};
\addlegendentry{$\psi=-5$dB}
\coordinate (spypoint) at (axis cs:4,0.42);
\coordinate (spyviewer) at (axis cs:4.,0.395);
\spy[width=2.5cm,height=1.3cm, every spy on node/.append style={name=c1}] on (spypoint) in node [fill=white] at (spyviewer);

\end{axis}

\begin{axis}[%
width=0.78\linewidth,
height=2.5cm,
at={(0.58in,1.2in)},
scale only axis,
xmin=0.08,
xmax=5,
xlabel style={font=\color{white!15!black}},
xlabel={$\gamma$},
ymin=0.15,,
ymax=0.167,
ytick={0.15,0.155,0.16,0.165},
yticklabels={0.15,0.155,0.16,0.165},
ylabel style={font=\color{white!15!black}},
ylabel={CRB $\theta_2 [\circ]$},
axis background/.style={fill=white},
xmajorgrids,
ymajorgrids,
legend columns=1,
legend style={at={(0,1)}, anchor=north west, legend cell align=left, align=left, draw=white!15!black,font=\small}
]
\addplot [color=mycolor1, line width=0.5pt]
  table[row sep=crcr]{%
0.1	0.151185090743883\\
1	0.153401937142034\\
2	0.156267495153226\\
4	0.162852628833013\\
5	0.166535134587396\\
};
\addlegendentry{$\psi=0$dB}

\addplot [color=mycolor2, line width=0.5pt]
  table[row sep=crcr]{%
0.1	0.151325548918935\\
1	0.153441691630772\\
2	0.156391843059308\\
4	0.16306900507959\\
5	0.166803221206146\\
};
\addlegendentry{$\psi=-3$dB}

\addplot [color=mycolor3, line width=0.5pt]
  table[row sep=crcr]{%
0.1	0.151169591514157\\
1	0.153672814024996\\
2	0.156558006376278\\
4	0.16322552760919\\
5	0.166987486334249\\
};
\addlegendentry{$\psi=-5$dB}

\coordinate (spypoint2) at (axis cs:4,0.163);
\coordinate (spyviewer2) at (axis cs:4,0.155);
\spy[width=2.3cm,height=1.3cm, every spy on node/.append style={name=a1}] on (spypoint2) in node [fill=white] at (spyviewer2);

\end{axis}
\end{tikzpicture}%

%% file: images/SNR_vs_psi/SNR_vs_psi.tex
\definecolor{mycolor1}{rgb}{0.00000,0.44700,0.74100}%
\definecolor{mycolor2}{rgb}{0.85000,0.32500,0.09800}%
\definecolor{mycolor3}{rgb}{0.92900,0.69400,0.12500}%
\begin{tikzpicture}

\begin{axis}[%
width=0.85\linewidth,
height=2cm,
at={(0.758in,0.582in)},
scale only axis,
xmin=0,
xmax=5,
xlabel style={font=\color{white!15!black}},
xlabel={$\gamma$},
ymin=-10,
ymax=10,
ylabel style={font=\color{white!15!black}},
ylabel={SNR [dB]},
axis background/.style={fill=white},
xmajorgrids,
ymajorgrids,
legend columns=3,
legend style={at={(0.03,1)}, anchor=north west, legend cell align=left, align=left, draw=white!15!black,font=\small}
]
\addplot [color=mycolor1, line width=1.0pt]
  table[row sep=crcr]{%
0.1	-0.0291442031072372\\
1	-0.00362527676348835\\
2	-0.00114755129187854\\
4	-0.00132841970830988\\
5	-0.000405761267903202\\
};
\addlegendentry{$\psi=0$dB}

\addplot [color=mycolor2, line width=1.0pt]
  table[row sep=crcr]{%
0.1	-3.03167058037143\\
1	-3.01372334904647\\
2	-3.01084443005366\\
4	-3.0119115959275\\
5	-3.01123887604834\\
};
\addlegendentry{$\psi=-3$dB}

\addplot [color=mycolor3, line width=1.0pt]
  table[row sep=crcr]{%
0.1	-5.23325185725648\\
1	-5.22963277664363\\
2	-5.231544876047\\
4	-5.22960707840996\\
5	-5.2292911513567\\
};
\addlegendentry{$\psi=-5$dB}

\end{axis}
\end{tikzpicture}%

%% file: images/close_vs_distant/CloseVSDistantSubplot.tex
\definecolor{mycolor1}{rgb}{0.00000,0.44700,0.74100}%
\definecolor{mycolor2}{rgb}{0.85000,0.32500,0.09800}%
\definecolor{mycolor3}{rgb}{0.92941,0.69412,0.12549}%
\definecolor{mycolor4}{rgb}{0.49412,0.18431,0.55686}%

\begin{tikzpicture}[baseline=(current bounding box.west)]

\begin{axis}[%
width=0.85\linewidth,
height=2cm,
at={(0.58in,2.54in)},
scale only axis,
xmin=0,
xmax=5,
xlabel={$\gamma$},
ymin=0.278,
ymax=0.82,
ylabel={$[\circ]$},
axis background/.style={fill=white},
xmajorgrids,
ymajorgrids,
legend columns=2,
legend style={at={(0,0.97)}, anchor=north west, legend cell align=left, align=left, draw=white!15!black,font=\small}
]
\addplot [color=mycolor1, line width=1.0pt,dashed]
  table[row sep=crcr]{%
0.1	0.389089587346206\\
1	0.400579804689153\\
2	0.437243223152719\\
4	0.597804482258471\\
5	0.82379729967463\\
};
\addlegendentry{CRB $\theta_1$ close}

\addplot [color=mycolor1, line width=1.0pt]
  table[row sep=crcr]{%
0.1	0.389312366886668\\
1	0.395034965578866\\
2	0.402433237970152\\
4	0.41942681677198\\
5	0.42892500943948\\
};
\addlegendentry{CRB $\theta_1$ }

\end{axis}

\begin{axis}[%
width=0.85\linewidth,
height=2cm,
at={(0.58in,1.3in)},
scale only axis,
xmin=0,
xmax=5,
xlabel={$\gamma$},
ymin=0.148,
ymax=0.34,
ylabel={$[\circ]$},
axis background/.style={fill=white},
xmajorgrids,
ymajorgrids,
legend columns=2,
legend style={at={(0,0.97)}, anchor=north west, legend cell align=left, align=left, draw=white!15!black,font=\small}
]
\addplot [color=mycolor2, line width=1.0pt,dashed]
  table[row sep=crcr]{%
0.1	0.151098926468545\\
1	0.155670627439026\\
2	0.170593213792519\\
4	0.236248643474264\\
5	0.331986587913232\\
};
\addlegendentry{CRB $\theta_2$ close}

\addplot [color=mycolor2, line width=1.0pt]
  table[row sep=crcr]{%
0.1	0.151185090743883\\
1	0.153401937142034\\
2	0.156267495153226\\
4	0.162852628833013\\
5	0.166535134587396\\
};
\addlegendentry{CRB $\theta_2$ }

\end{axis}
\end{tikzpicture}%